\documentstyle[epsfig]{article}

\def\tr{{\rm tr}}
\begin{document}
\centerline{NON-LOCALITY OF QUANTUM MECHANICS}
\centerline{AND THE LOCALITY OF GENERAL RELATIVITY}
\bigskip
\centerline {Jeeva Anandan$^*$}
\smallskip
\centerline{Clarendon Laboratory}
\centerline{University of Oxford}
\centerline{Parks Road}
\centerline{Oxford, ~~UK}

\centerline{and}

\centerline {School of Natural Sciences }
\centerline {Institute for Advanced Study}
\centerline {Princeton, NJ 08540, USA }

\centerline{E-mail: jeeva@sc.edu}

\bigskip

\begin{abstract}
The conflict between the locality of general relativity, reflected
in its space-time description, and the non-locality of quantum
mechanics, contained in its Hilbert space description, is
discussed. Gauge covariant non-local observables that depend on gauge
fields and gravity as well as the wave function are used in order
to try to understand and minimize this conflict within the frame-work of these
two theories. Applications are made to the
Aharonov-Bohm effect and its generalizations to non Abelian gauge
fields and gravity.
\end{abstract}

\bigskip\noindent
{\it Invited talk delivered at the Meeting on the Interface of
Gravitational and Quantum Realms, IUCAA, Pune, India, Dec. 17-21,
2001. \\To be published in Modern Physics Letters A.}

\bigskip
\bigskip\noindent
*On leave from the Department of Physics and Astronomy, University
of South Carolina, Columbia, SC 29208, USA.
\bigskip

\noindent{May 13, 2002. Revised June 10, 2002.}

\newpage
\section{Introduction}

At present, the two most fundamental theories in physics are quantum theory and
general relativity, which is a classical theory of the gravitational field.
These two theories are incompatible. A major problem therefore is to unify these
two theories in the sense that they are approximations of a deeper quantum
theory of gravity. Perhaps the deepest level at which the incompatibility
between the two theories is manifest is in the fact that general relativity is
local whereas quantum mechanics is non-local. This is reflected in the
formulation of general relativity geometrically in {\it space-time}, whereas
quantum mechanics is formulated in {\it Hilbert spaces} whose elements (state
vectors or wave functions) have non local properties in space-time.

This fundamental incompatibility will be described in more detail in section 2.
The translation group elements, which are non local observables, are used to
represent the non-local aspects of the quantum wave function, as shown for
example in interference phenomena, through their expectation values. In section
3, the non local quantum phenomena, such as the Aharonov-Bohm (AB) effect and it
generalization to non-Abelian gauge fields, are described by means of gauge
invariant non local observables that are elements of the direct product of the
translation group and the gauge group, following earlier work of Aharonov and
the author. The gravitational AB effect around a cosmic string is considered in
section 4. It is shown that this effect may be described as the change in the
expectation value of a Poincare group element. This naturally leads to a
description of the gravitational field by means of the Poincare group.

\section{Locality versus Non-Locality}

In a curved space-time, owing to the locality of all the interactions, it is not
meaningful {\it a priori} to compare two tangent vectors $v$ and $v'$ at
different space-time points $p$ and $p'$.
But $v'$ may be parallel transported with respect to a {\it connection} along a
curve joining $p'$ and $p$ and the resulting vector $v''$ can be compared with
$v$.
The different such vectors $v''$ corresponding to different choices of the curve
are related by holonomy transformations (parallel transport around a closed
curve). Thus ironically, the locality of general relativity leads us to consider
holonomy transformations which appear to be non locally associated with an
entire closed curve in space-time.

Similarly, in the presence of a gauge field, owing to the locality of the gauge
field interaction, it is not possible to compare two internal vectors (which
belong to a representation of the gauge group) at different points in space-
time.
The holonomy transformation in the case of the electromagnetic gauge field is
\begin{equation}
u = \exp( -i{\hat q\over \hbar c} \oint A_\mu dy^\mu)
\label{holonomy}
\end{equation}
where $ A_\mu$ is the $4-$vector potential that acts on a wave function and
$\hat q$ is the charge operator. There is, however, a fundamental difference
between the non-locality represented by (\ref{holonomy}) and the apparent non-
locality of the holonomy transformation in a gravitational field. This is due to
the soldering form, known to mathematicians as the canonical $1-$form, in a
gravitational field, which `solders' the parallel transportable vectors to the
tangent space at each point on space-time \cite{tr1973}. Consequently, it is
possible to relate the orientation of the vector that is parallel transported
along a curve with the orientation of the tangent vector to the curve at each
point on the curve, which cannot be done for the parallel transport of an
`internal' vector by means of the gauge field connection. In particular, while
the holonomy transformation is unchanged if the enclosed `flux' is changed by
one `quantum', the two situations may be distinguished in the gravitational case
by the above mentioned comparison with the tangent vector, but they cannot be
distinguished in the electromagnetic or gauge field case \cite{an1993}. This
fundamental difference exists between gravitational and gauge fields even when
the vector parallel transported is classical, because the soldering form is part
of the gravitational field. But there is a further quantum non-locality that
exists in the non-local nature of the wave function in both gravity and gauge
fields.

This quantum non-locality may be seen already in the simple
double slit experiment.  Suppose the two slits are separated by a
distance $l$ in the y-direction and the quantum wave is moving in
the x-direction that is normal to the two slits. Soon after the
wave passes the double slit, it is a superposition of two wave
packets at the two slits:
\begin{equation}
\psi(x,y,z,t)= {1\over\sqrt{2}}\{ \phi(x,y-\ell,z,t)+ e^{i\alpha}
\phi(x,y,z,t)\}
\label{wave}
\end{equation}
For simplicity, it is assumed here that the two wave packets are the same except
for the phase difference $\alpha$. However, {\it no local experiments done on
the non overlapping wave packets emerging from the slits could determine this
phase difference}. For example, the
expectation  values
of the local variables $p^n$, where $n$ is any positive
integer, give no
information about $ e^{i\alpha}$. (This is easily verified from
$p^n= (-i\hbar {\partial\over \partial y})^n$ in the
coordinate representation.)
But
\begin{equation}
<\psi|\exp(-i {p \ell\over \hbar})|\psi> = {e^{i\alpha} \over 2}
\label{alpha}
\end{equation}
This $e^{i\alpha}$ is of course observed when the two wave packets overlap
subsequently to produce inerference fringes. This justifies treating the
translational group element $s=\exp(-i{pl\over \hbar})$ as an observable, and $s^\dagger$
was called modular momentum by Aharonov et al \cite{ah1969,hermitian}. The
reason why $<\psi|\exp(-i {p \ell\over \hbar})|\psi>$ contains more information
than all the $<\psi|p^n|\psi>$ is because $|\psi>$ vanishes in a region between
the wave packets and therefore $<\psi| \exp(-i {p \ell\over \hbar})|\psi>$ is
not analytic. However, the latter is well defined through the action of $\exp(-i
{p \ell\over \hbar})$ on the momentum space wave function of $|\psi>$ by
multiplication.

A great importance of the observable $s$ lies in the fact that it
determines whether the state is a pure state or a mixture.
Suppose we have a mixture of the above mentioned localized wave
packets, denoted $|\psi_1>$ and $|\psi_2>$, instead of the above
pure state $|\psi>= {1\over\sqrt{2}}(|\psi_1> + |\psi_2>)$. The
density matrix of this mixture is $\rho = {1\over 2}(|\psi_1>
<\psi_1|+ |\psi_2><\psi_2|)$. Now $\rho$ gives the same results
for the observation of all local observables as the density
matrix $\rho_\psi= |\psi><\psi|$ of the pure state, i.e.
\begin{equation}
\tr
(\rho \hat A) = \tr (\rho_\psi \hat A)={1\over 2}( <\psi_1|\hat
A|\psi_1>+<\psi_2|\hat A|\psi_2>)
\end{equation}
for every local observable
$\hat A$. However, $\tr(\rho s) =0$, whereas $\tr (\rho_\psi s) =
{e^{i\alpha} \over 2}$, which is the same as eqn. (\ref{alpha})
above. Here $\tr (\rho_\psi s)\ne 0$ because $\rho_\psi$ is the
density matrix of a coherent superposition of the two wave
packets unlike $\rho$. Hence $s$ may be used to represent the
quantum coherence between the two wave packets.

This remark becomes important when we probe the gravitational
field by means of quantum mechanical particles, instead of the
classical particles that Einstein used to obtain his principle of
equivalence. If we probe the classical gravitational field using
a particle whose state is $|\psi>$, and measure $|\psi>$ by means
of local observables, this changes it into a {\it mixture} or is
indistinguishable from a mixture. This is because the interaction
of a classical gravitational field is local, which enables it to
be incorporated into the space-time geometry. In this way we
determine the {\it classical} gravitational field and the
associated space-time geometry. We need a non local observable,
such as the translational group element $s=\exp(-ip\ell)$ in
order to probe the quantum coherence of the gravitational field.
Also, the backreaction of $|\psi>$ that represents the superposed
wave packets requires quantum gravity, which should be non-local
because of the above non-local aspect of $|\psi>$. Hence,
although non-local operators cannot be observed by present day
experiments because of the locality of the interactions, their
study may be useful in understanding the quantum aspects of the
gravitational field. This requires that $s$ should be generalized
to incorporate the change in space-time geometry due to the
gravitational field which will be attempted in section 4. But in
the next section we shall consider the generalization of $s$ to
gauge fields, which are easier than the gravitational field
because they do not require modification of the Minkowski
space-time and it associated symmetries.

\section{Non Locality of the Aharonov-Bohm Effect}

The Aharonov-Bohm (AB) effect \cite{ah1959}, which is a non local interaction
between the wave function and the electromagnetic field, gives important
information about both the field and the wave function. It shows, on the one
hand, that the complete description of the electromagnetic field are given by
holonomy transformations (\ref{holonomy}) associated with arbitrary closed
curves. The field strength $F_{\mu\nu}$ has too little information in a multiply
connected region, while the phase ${q\over \hbar c} \oint A_\mu dy^\mu$ has too
much information about the electromagnetic field because it may be changed by
$2\pi n$ ($n$ is an integer), without observable consequences \cite{wu1975}, as
mentioned in the previous section. Since (\ref{holonomy}) is an element of the
$U(1)$ group, the AB effect is telling us that the electromagnetic field is a
$U(1)$ gauge field. On the other hand, the AB effect also shows the non local
quantum coherence of the wave function discussed in the previous section. It is
therefore natural to try to describe this effect using quantities that depend on
both the wave function and the field, as will be done shortly.

The usual way of explaining the AB effect is by means of the gauge invariant
holonomy transformation (\ref{holonomy}) associated with {\it closed curves},
which depends only on the external field. However, gauge invariant quantities
associated with {\it open curves} that depend on both the field and the wave
function were used previously to study the AB effect in superconductors
\cite{an1986}. An example of such a quantity for a particle with charge $q$ at a
given time $t$ is \cite{an1986}
\begin{equation}
z({\bf x},t,\mbox{\boldmath $\ell$})= \psi^*({\bf x}+{\mbox{\boldmath
$\ell$}},t)
\exp \{i{q\over \hbar} \int_{\bf x}^{{\bf x}+{\mbox{\boldmath $\ell$}}} {\bf
A}({\bf y},t)\cdot
d{\bf y}\} \psi({\bf x},t)
\label{z}
\end{equation}
It is easy to verify that (\ref{z}) is invariant under the gauge transformation
$$\psi'(x)= \exp\{i{q\over \hbar}
\Lambda(x)\}\psi(x), ~~{\bf A}'(x) = {\bf A}(x)+\nabla \Lambda(x)$$
where $x$ stands for $({\bf x}, t)$, which will also be denoted by $x^\mu$. At
present, for simplicity, the integral in (\ref{z}) is taken along the straight
line
joining $({\bf x},t)$ and $({\bf x}+{\mbox{\boldmath $\ell$}},t)$, although the
gauge invariance of $z({\bf x},t,\mbox{\boldmath $\ell$})$ holds more generally
for integral along any piece-wise differentiable curve joining these two points.

To construct an observable corresponding to the gauge invariant quantities $z$,
consider the tranformation
$\psi_{\mbox{\boldmath $\ell$}}'(x)= f_{\mbox{\boldmath $\ell$}} (x) \psi(x)$,
where \cite{ah}
\begin{equation}
f_{\mbox{\boldmath $\ell$}}(x) = \exp (-{i\over\hbar}\hat{\bf p}\cdot
{\mbox{\boldmath $\ell$}}) \exp \{i{q\over \hbar} \int_{\bf x}^{{\bf
x}+{\mbox{\boldmath $\ell$}}} {\bf A}({\bf y},t)\cdot
d{\bf y}\}
\label{f}
\end{equation}
It follows that
\begin{eqnarray}
<\psi|f_{\mbox{\boldmath $\ell$}}|\psi>&=&<\exp ({i\over\hbar}{\bf p}\cdot
{\mbox{\boldmath $\ell$}})\psi|
\exp \{i{q\over \hbar} \int_{\bf x}^{{\bf x}+{\mbox{\boldmath $\ell$}}} {\bf
A}({\bf y},t)\cdot
d{\bf y} \}|\psi>\nonumber \\
&=&\int d^3x \psi^*({\bf x}+{\mbox{\boldmath $\ell$}},t) \exp
\{i{q\over \hbar} \int_{\bf x}^{{\bf x}+{\mbox{\boldmath
$\ell$}}} {\bf A}({\bf y},t)\cdot d{\bf y}\} \psi({\bf x},t)
\label{<f>}
\end{eqnarray}
The integrand of the last expression is the same as the $z$ given
by (\ref{z}), which is therefore more general than (\ref{<f>}).
Since (\ref{z}) is gauge invariant, so is (\ref{f}). Therefore,
(\ref{f}) is gauge covariant.

Consider now the translational group element $\exp(-{i\over \hbar}{\bf p}\cdot
{\mbox{\boldmath $\ell$}})$. This operator determines the quantum coherence of a
wave function, as discussed in the previous section. This is shown dramatically
in the AB effect \cite{ah1959} due to interference of two wave packets around a
solenoid. It was shown by Aharonov et al \cite{ah1969} that while there is no
exchange of ${\bf p}^n$ for any positive integer $n$, there is an exchange of
$\exp(ip\ell)$ between the solenoid and the electron. To see this, consider a
wave that is a superposition of two localized wave packets, as in (\ref{wave}),
which pass on opposite sides of the solenoid, with $\alpha$ now being the AB
phase caused by the solenoid. Now, while $<\psi|{\bf p}^n|\psi>$ is independent
of $\alpha$, $<\psi| \exp(-{i\over \hbar}{\bf p}\cdot {\mbox{\boldmath
$\ell$}})|\psi>$ depends on $\alpha$ for appropriate choice of ${\mbox{\boldmath
$\ell$}}$ as shown by (\ref{alpha}). Therefore, while $<\psi|{\bf p}^n|\psi>$ is
unaffected by the AB phase, $<\psi| \exp(-{i\over \hbar}{\bf p}\cdot
{\mbox{\boldmath $\ell$}})|\psi>$ is changed by it. This difference is due to
the non locality of the quantum wave which, as mentioned in the previous
section, exists due to the quantum coherence of the two wave packets.

But $\alpha$ in the above analysis is gauge dependent. Consider
therefore the operator $ {f'}_{\mbox{\boldmath $\ell$}}=\exp
\{i(-{\bf p}+q\hat{\bf A}(t)) \cdot {\mbox{\boldmath $\ell$}}\}$ that
is obtained by replacing the canonical momentum $\bf p$ in the
translation with the gauge-covariant kinetic momentum ${\bf
p}-q\hat{\bf A}(t)$, where $\hat{\bf A}(t)$ is a time-dependent
operator in the Schr\"odinger picture that acts on the Hilbert
space according to the rule
\begin{equation}
\hat{\bf A}(t)\psi({\bf x})= {\bf A}({\bf x},t) \psi({\bf x})
\end{equation}
Here $\psi({\bf x})$ is any element of the Hilbert space of the particle and
${\bf A}( {\bf x},t)$ is the usual classical vector potential in space-time. If
the electromagnetic field is also quantized then ${\bf A}( {\bf x},t)$ would be
the expectation value of a time-independent Schr\"odinger operator representing
the quantized field with respect to a Schr\"odinger coherent state of the field
at time $t$. For this reason, I shall always let $ {f'}_{\mbox{\boldmath
$\ell$}}$ act on the Schr\"odinger wave function of the particle at time $t$,
because the latter wave function multiplies the coherent state wave function of
the field at the same time in the present approximation. Thus this restriction
may be deduced from a deeper theory in which all degrees of freedom are
quantized. Then the wave function resulting from the action of $
{f'}_{\mbox{\boldmath $\ell$}}$ is gauge-covariant owing to the fact that $-{\bf
p}+q\hat{\bf A}(t)$ is gauge-covariant.

Also, $f'_{\mbox{\boldmath $\ell$}} = f_{\mbox{\boldmath
$\ell$}}$ \cite{ah}, i.e. for every wave function $\psi$,
\begin{equation}
\exp \{{i\over\hbar}(-{\bf p}+q\hat{\bf A}) \cdot {\mbox{\boldmath
$\ell$}}\}\psi(x) = \exp (-{i\over\hbar}{\bf p} \cdot
{\mbox{\boldmath $\ell$}})\exp \{i{q\over \hbar} \int_{\bf
x}^{{\bf x}+{\mbox{\boldmath $\ell$}}} {\bf A}({\bf y},t)\cdot
d{\bf y}\}\psi(x) \label{f'}
\end{equation}
To prove this, note that (\ref{f'}) is valid in an axial gauge in which
$\hat{\bf A}(t) \cdot {\mbox{\boldmath $\ell$}}=0$. Also, both sides of (\ref{f'})
are gauge-covariant, as shown earlier. Hence, (\ref{f'}) is valid in every
gauge.

More generally, define the operator $g_{\hat\gamma}$ that acts on the Hilbert
space
according to \cite{an2002}
\begin{equation}
g_{\hat\gamma}\psi(x) = \exp (-{i\over\hbar}{\bf p} \cdot {\mbox{\boldmath
$\ell$}}) \exp \{i{q\over \hbar} \int\limits_{~\gamma~\bf x}^{{\bf
x}+{\mbox{\boldmath $\ell$}}} {\bf A}({\bf y},t)\cdot d{\bf
y}\}\psi(x)
\label{gamma}
\end{equation}
where $\hat\gamma$ is {\it any} piece-wise differentiable curve in
${\cal R}^3$ and $\gamma$ is a curve in physical space that is
congruent to $\hat\gamma$ and joins an arbitrary point $\bf x$ to
${\bf x}+{\mbox{\boldmath $\ell$}}$. Here, ${\mbox{\boldmath
$\ell$}}$ is independent of $\bf x$ but depends on $\hat\gamma$.
In the special case when $\hat\gamma$ is a straight line,
$g_{\hat\gamma}$ is the same as $ f_{\mbox{\boldmath $\ell$}}$.
Also, for any piece-wise differentiable curve $\hat\gamma$,
$g_{\hat\gamma}$ can be shown to  be gauge-covariant in the same
way as $ f_{\mbox{\boldmath $\ell$}}$ was shown to be
gauge-covariant.
From now onwards, the $\hat{}$ over $\gamma$ will
be dropped, for simplicity, if no confusion will arise.

Consider now the AB effect \cite{ah1959} in which a wave packet is split by a
beam splitter into two wave packets that go past on two sides of a solenoid
(figure 1).

%\centerline{\psfig{figure=abcurve.eps,height=3.5in}}
\centerline{\psfig{figure=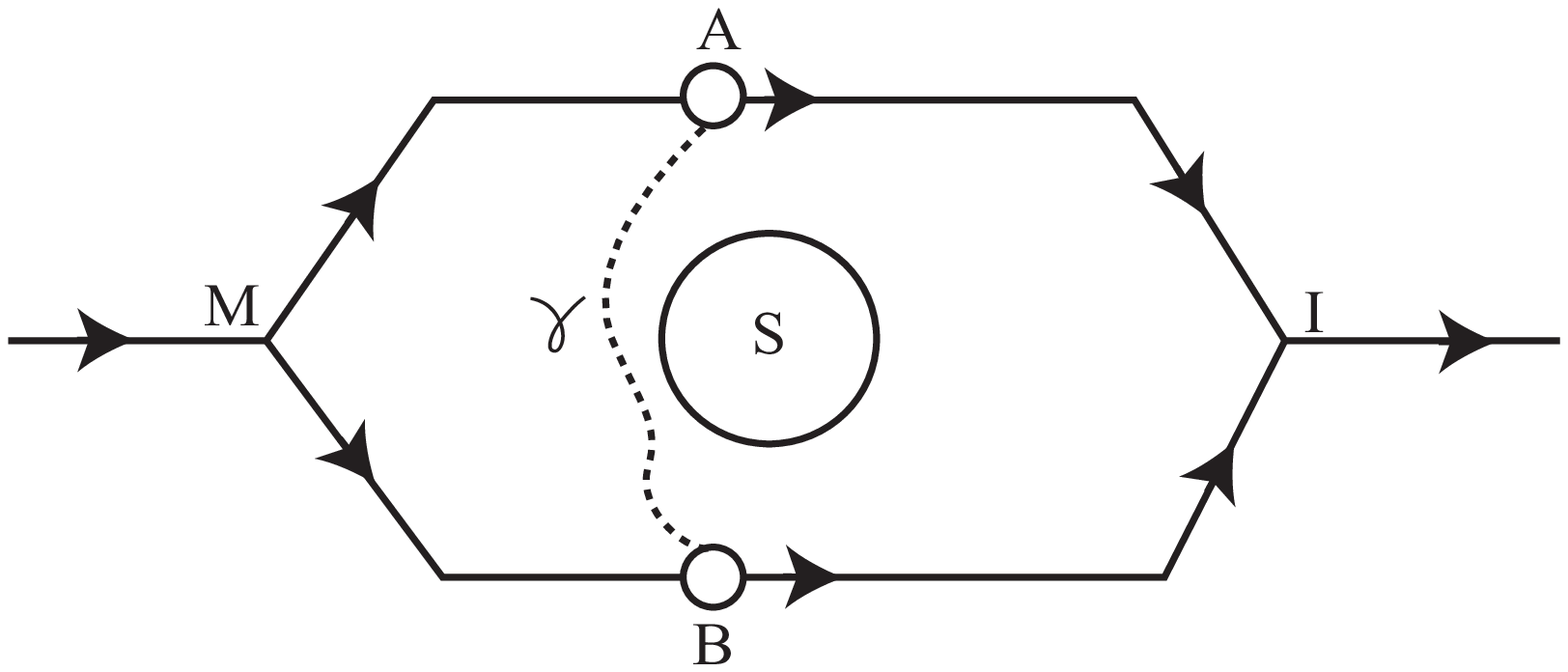,height=2in}}

\bigskip

\noindent
Figure 1. A wave packet is split at the beam splitter M into two wave
packets A and B that go around the solenoid S and interfere at I. When
the imaginary curve $\gamma$ crosses the solenoid, the {\it gauge
invariant} expectation value of $g_\gamma$ changes.

\bigskip

The wave function of the particle is
\begin{equation}
\psi( {\bf x},t)= {1\over\sqrt{2}}\{ \psi_1({\bf x},t)+
\psi_2({\bf x},t)\}
\label{psi}
\end{equation}
where $\psi_1$ and $\psi_2$ are the wave functions of
the two wave packets. Then we may write
\begin{eqnarray}
\psi_1({\bf x},t)&=& \exp \{i{q\over \hbar}
\int\limits_{~\gamma_1~{\bf x}_0}^{\bf x} {\bf A}({\bf y},t)\cdot
d{\bf y}\}\psi_{10}({\bf x},t), \nonumber\\
\psi_2({\bf x},t)&=& \exp
\{i{q\over \hbar} \int\limits_{~\gamma_2~{\bf x}_0}^{\bf x} {\bf
A}({\bf y},t)\cdot d{\bf y}\}\psi_{20}({\bf x},t)
\label{wavepackets}
\end{eqnarray}
where $\gamma_1$ and $\gamma_2$ are variable curves outside the solenoid that
begin at a point ${\bf x}_0$ on the beam splitter, and $\psi_{10}$ and
$\psi_{20}$ are the unperturbed solutions of Schr\"odinger's equation in the
absence of $\bf A$. It is easily verified that $\psi_1$ and $\psi_2$ are
solutions of Schr\"odinger's equation with $\bf A$ minimally coupled. Suppose
the displacement between the centers of the wave packets is $\bf\ell$. Then
\begin{eqnarray}
<\psi|g_\gamma|\psi>&=&<\exp ({i\over\hbar}{\bf p}\cdot {\mbox{\boldmath
$\ell$}})\psi|
\exp \{i{q\over \hbar} \int\limits_{~\gamma~{\bf x}}^{{\bf x}+{\mbox{\boldmath
$\ell$}}} {\bf A}({\bf y},t)\cdot
d{\bf y} \}|\psi>\nonumber \\
&=& \int d^3x \psi^*({\bf x}+{\mbox{\boldmath $\ell$}},t)
\exp \{i{q\over \hbar} \int\limits_{~\gamma~{\bf x}}^{{\bf x}+{\mbox{\boldmath
$\ell$}}}{\bf A}({\bf y},t)\cdot
d{\bf y}\} \psi({\bf x},t)\nonumber \\
&=&\int d^3x \psi_1^*({\bf x}+{\mbox{\boldmath $\ell$}},t)
\exp \{i{q\over \hbar} \int\limits_{~\gamma~{\bf x}}^{{\bf x}+{\mbox{\boldmath
$\ell$}}}{\bf A}({\bf y},t)\cdot
d{\bf y}\} \psi_2({\bf x},t)\nonumber \\&=&
\int d^3x \psi_{10}^*({\bf x}+{\mbox{\boldmath $\ell$}},t)
\exp \{i{q\over \hbar} \oint_{\gamma_0}{\bf A}({\bf y},t)\cdot
d{\bf y}\} \psi_{20}({\bf x},t)
\label{<ab>}
\end{eqnarray}
on using (\ref{wavepackets}), where $\gamma_0$ is the closed curve formed
by $\gamma_2, \gamma$ and the reverse of $\gamma_1$.

As the two wave packets move past the solenoid, $\gamma$ crosses the solenoid.
Then, $\exp \{i{q\over \hbar} \oint_{\gamma_0} {\bf A}({\bf y},t)\cdot
d{\bf y}\}$ changes from $1$ to $\exp(i{q\over \hbar}\Phi)$, where $\Phi$ is
the magnetic flux enclosed by the solenoid. This has been shown for the special
case when $\gamma$ is a straight line joining the wave packets in a special
gauge \cite{ah2002}. But the present gauge invariant treatment which is valid
for arbitrary $\gamma$ has the following advantages: A) There is nothing special
about the time when the imaginary straight line joining the wave packets crosses
the solenoid, because for other choices of the curve $\gamma$, this curve would
cross the solenoid at different times. B) Even at a given time $t$, $\gamma$ may
be chosen to be on either side of the magnetic flux, which correspond to $\exp
\{i{q\over \hbar} \oint_{\gamma_0} {\bf A}({\bf y},t)\cdot
d{\bf y}\} =1 $ or $\exp(i{q\over \hbar}\Phi)$. Consider a circular
superconducting ring enclosing the magnetic flux $\Phi$ that has
${\mbox{\boldmath $\ell$}})$ as a diameter. The ring is interrupted by a
Josphson junction on one side of this diameter. Then $\gamma$ may be chosen to
be through the semi-circular ring on either side of this diameter. For these two
choices, $<\psi|g_\gamma|\psi>$ would take two possible values that differ by
the factor $\exp(i{q\over \hbar}\Phi)$. This difference is responsible for the
Josephson current through the Josephson junction. To summarize, the usual AB
effect described in fig. 1 as well as the AB effect in a superconducting ring
may be understood as due to the change in $<\psi|g_\gamma|\psi>$ that depends on
the open curve $\gamma$. But this change depends on the holonomy transformation
(\ref{holonomy}) associated with a closed curve.

\section{Generalization to Non-Abelian Gauge Fields and the
Gravitational Aharonov-Bohm Effect}

The results in the previous section naturally generalize to non
Abelian gauge fields \cite{an2002}. For an arbitrary gauge field,
(\ref{gamma}) is generalized to
\begin{equation}
g_{\hat\gamma}\psi(x) = \exp (-{i\over\hbar}{\bf p} \cdot {\mbox{\boldmath
$\ell$}}) P\exp \{i{g_0\over \hbar} \int\limits_{~\gamma~\bf x}^{{\bf
x}+{\mbox{\boldmath $\ell$}}} T_k{\bf A}^k({\bf y},t)\cdot d{\bf
y}\}\psi(x)
\label{gammagauge}
\end{equation}
where $T_k$ generate the gauge group, and $P$ denotes
path ordering, which is necessary because
$T_k{\bf A}^k $ at different points do not commute in general. In the special
case when $\gamma$ is a straight line,
\begin{equation}
\exp \{{i\over \hbar}(-{\bf p}+g_0 T_k\hat{\bf A}^k) \cdot {\mbox{\boldmath
$\ell$}}\}\psi(x) = \exp (-{i\over\hbar}{\bf p} \cdot {\mbox{\boldmath
$\ell$}})\exp \{i{g_0\over \hbar} \int_{\bf x}^{{\bf
x}+{\mbox{\boldmath $\ell$}}}T_k {\bf A}^k({\bf y},t)\cdot d{\bf
y}\}\psi(x)
\label{g'}
\end{equation}
This result may be proved, as in the Abelian case, using the gauge
covariance of both sides of (\ref{g'}) and noting that (\ref{g'})
holds in the axial gauge in which there is no contribution from
the non Abelian vector potential. Also, this result can be
generalized to an arbitrary (piece-wise differentiable) curve
$\hat\gamma$, i.e.
\begin{equation}
g_{\hat\gamma}\psi(x) = P\exp \{{i\over
\hbar}\int_{\hat\gamma}(-{\bf p}+g_0 T_k\hat{\bf A}^k) \cdot
d{\mbox{\boldmath $\ell$}}\} \label{group}
\end{equation}
where $P$ denotes path ordering. The result (\ref{group}) implies
that the operators $\{g_{\hat\gamma}\}$ form a group, which in
general is infinite dimensional.

Also, the AB effect described in fig. 1 may be generalized to the case when the
flux through the cylinder is due to an arbitrary gauge field. The wave function
is given by (\ref{psi}), where now
\begin{eqnarray}
\psi_1({\bf x},t)&=& P\exp \{i{g_0\over \hbar}
\int\limits_{~\gamma_1~{\bf x}_0}^{\bf x} T_k{\bf A}^k({\bf y},t)\cdot
d{\bf y}\}\psi_{10}({\bf x},t), \nonumber\\
\psi_2({\bf x},t)&=& P\exp
\{i{g_0\over \hbar} \int\limits_{~\gamma_2~{\bf x}_0}^{\bf x}
T_k{\bf A}^k({\bf y},t)\cdot d{\bf y}\}\psi_{20}({\bf x},t)
\label{napackets}
\end{eqnarray}
Then,
\begin{equation}
<\psi|g_\gamma|\psi>=\int d^3x \psi_{10}^\dagger({\bf x}+{\mbox{\boldmath
$\ell$}},t)
P\exp \{i{g_0\over \hbar} \oint_{\gamma_0} T_k{\bf A}^k({\bf y},t)\cdot
d{\bf y}\} \psi_{20}({\bf x},t)
\end{equation}
where $\gamma_0$ is the closed curve, as defined before, beginning and ending at
${\bf x}_0$. The gauge invariance of this expression follows from the fact
that in (\ref{napackets}) the parallel transport operators (path ordered
integrals) begin at ${\bf  x}_0$. Therefore, under a
local gauge transformation $U({\bf  x},t)$, the unperturbed wave
functions $\psi_{10}(x)$ and $\psi_{20}(x)$ must undergo the global
gauge transformation $U({\bf  x}_0,t)$ in order that $\psi_1(x)$ and $\psi_2(x)$
transform covariantly to $U(x)\psi_1(x)$ and $U(x)\psi_2(x)$, respectively. The
AB effect in the
superconducting ring may be generalized to the AB effect due to
an arbitrary gauge field flux that arises from the generalization
of the Josephson effect to a non Abelian gauge theory
\cite{an1986}. In both these cases, as discussed in the previous
section for the $U(1)$ gauge theory, the generalized AB effect
arises from the change in $<\psi|g_\gamma|\psi>$ due to the crossing of the flux
by $\gamma$, where now $g_\gamma$ is given by
(\ref{gammagauge}).

Extending these results to the gravitational case, however, is difficult because
of the fundamental locality of the gravitational field due to space-time
curvature \cite{an1993}. To this end, consider the gravitational AB effect due
to a cosmic string. In particular, in the experiment described in fig. 1, the
solenoid may be replaced by a cosmic string. Outside the cosmic string, the
curvature and torsion vanish, which is analogous to the vanishing of the
electromagnetic field strength outside the solenoid. Define a Hilbert space
$\cal H$ to consist of $L^2$ functions that have support outside the cosmic
string. Now generalize $g_\gamma$ to include the gravitational field
\cite{an1980,an1996}:
\begin{equation}
g_\gamma =P\exp\{{i\over\hbar}\int_\gamma ({\theta_\mu}^a P_a
+{1\over2} {{{\omega}_\mu}^a}_b {M^b}_a +{A_\mu}^jT_j) dx^\mu\},
\end{equation}
where ${\theta_\mu}^a $ is the soldering form mentioned in section
2 , ${{{\omega}_\mu}^a}_b$ is the linear connection form, and
$P_a$ and ${M^b}_a$ generate the space-time translations and
Lorentz transformations in the Hilbert space $\cal H$. Hence,
\begin{equation}
<\psi|g_\gamma|\psi>=\int d^3x \psi_{10}^\dagger({\bf x}+{\mbox{\boldmath
$\ell$}},t)
g_{\gamma_0} \psi_{20}({\bf x},t)
\end{equation}
For simplicity, suppose that there is no gauge field present. It
was shown \cite{an1994} that the gravitational phase shift is
determined by $g_{\gamma_0}$ when $\gamma_0$ encloses the cosmic
string. The part of $g_{\gamma_0}$ that contributes to the phase
shifts are symmetries of the space-time geometry of the cosmic
string. So, again the change in $<\psi|g_\gamma|\psi>$ due to
$\gamma$ crossing the cosmic string (which makes $\gamma_0$ that
was defined below (\ref{<ab>}) enclose the cosmic string)
determines the gravitational AB effect in this case. The
generalization of the above results to an arbitrary gravitational
field will be discussed elsewhere.

%%%
\bigskip

\noindent
ACKNOWLEDGMENTS
\bigskip

I thank Todd Brun for a useful discussion. I also thank Stephen L. Adler for his hospitality at the
Institute for Advanced Study at Princeton. This work was
partially supported by a Fulbright award and the NSF Grant no.
9971005.

\end{document}